\documentstyle[12pt,qqaalart]{article}

\author{Jos\'e Antonio Belinch\'on\thanks{E-mail: abelinchon@caminos.recol.es}\\
Grupo Interuniversitario de An\'alisis Dimensional\\
Dept. de F\'isica ETS Arquitectura. UPM\\
Av. Juan de Herrea 4 28040 Madrid Espa\~na}
\title{Cosmological models with variable ``constants".
}
\date{September 1999
}
\input tcilatex

\QQQ{Language}{
American English
}

\begin{document}

\maketitle
\begin{abstract}
{\em The behavior of the ``constants'', $G,$ $c,$ $\hbar ,$ $a,$ $e,$ $m_i$
and $\Lambda ,$ considering them as variable, in the framework of a flat
cosmological model with FRW symmetries described by a bulk viscous fluid and
considering mechanisms of adiabatic matter creation are investigated. Two
cases are studied; one with radiation predominance and another of matter
predominance. It is found that with the solution obtained our model verifies
these basic principles: Lorentz invariance and general covariance, Mach,
Equivalence and causality. Finally to emphasize that the envisaged models
are free from the main problem: Planck's, horizon and entropy. With regard
to that model with matter predominance it is seen that mechanisms of
creation of matter cannot be considered since if these are taken into
account the temperature would increase instead of remaining constant while
the universe expands.}
\end{abstract}

\section{\bf Introduction.}

In a recent paper \cite{T0} the behavior of the ``constant'' $G,$ $c$ and $%
\Lambda $ was investigated within a model described by a bulk viscous fluid
and taking into account mechanisms of matter creation to solve the entropy
problem. Upon considering the constant $c$ as a function dependent on time $%
t,$ the condition that the radiation constant $a$ shoul be constant in the
same way that Boltzmann's constant $k_B$ was imposed. With this supposition
the following is obtained: Planck's constant $\hbar $ should behave as $%
\hbar \propto c^{-1}$. In this paper this point is taken up once more but
witout similar hypothesis i.e. Boltzmann's constant $k_B$ is the only
constant considered real. Therefore we suppose that all the ``constants'' $%
G, $ $c,$ $\hbar ,$ $a,$ $e,$ $m_i$ and $\Lambda $ are variable, without
making any previous hypothesis about their behavior or verifying any
equality in particular. Calculations are made within the framework described
above, regarding their behavior together with the rest of the quantities
wich characterize the model: $f,$ $\rho ,$ $\rho _m,$ $\theta ,$ $S,$ $s,$ $%
\xi $ and $n$ where respectively they represent the radius of the universe,
energy density, matter density, temperature, entropy, entropy density,
viscosity coefficient and particle number density.

Once calculated all these quantities, two concrete solutions are studied:
one that it would describe a universe with radiation predominance and
another with matter predominance, simplifying both solutions to the case of
non creation of matter i.e. $\beta =0.$ It is found, for example, that with
these solutions, it is always verified that $G/c^2$ (general covariance)
stays constant in both cases, independently of the value of $\beta $. The
expression $\rho =a\theta ^4$ is recovered for energy density. All energies
are preserved, but no the moment in the case of matter predominance, while
in the case of radiation predominance, the energy follows the law $E\propto
t^{-1/2}$ while the moment is constant. The fine structure constant $\alpha
, $ in both cases, continues being a real constant in spite of the fact that
all the constant that define it vary. The models described here verify the
following basic principles: as already indicated general covariance, it is
also shown that the principle of Lorentz invariance is verified, Mach,
Equivalence and causality. Both model lack the designated horizon problem
since the relationship: $f=ct$ is always verified. With the solutions
obtained, it is seen that the model lacks the designated problem of Planck
in the same way as that of entropy.

The paper is organized as follows: In the second section the governing
equations of our model are shown and considerations on the dimensional
method followed, are made. In the third section, use is made of the D.A. (Pi
theorem) to obtain a solution to the principal quantities that appears in
the model. Finally in the fourth section presentation is made of two
particular cases of the obtained solutions together with some conclusions.

\section{\bf The model.}

For a flat universe $k=0$ with FRW symmetries i.e. homogeneity and isotropy
were assumed and therefore there will be no spatial variations of
``constants'' $G,c$ and $\Lambda $ solely temporary. It is also supposed
that our fluid is bulk viscous (second viscosity) and mechanisms of creation
of matter are considered. With these suppositions the equations that govern
the model are as follows: 
\begin{equation}
\label{e1}2\frac{f\,^{\prime \prime }}{f\,}+\frac{(f\,^{\prime })^2}{f\,^2}=-
\frac{8\pi G(t)}{c^2(t)}(p+p_c)+c^2(t)\Lambda (t)\ \ 
\end{equation}

\begin{equation}
\label{e2}3\frac{(f\,^{\prime })^2}{f\,^2}=\frac{8\pi G(t)}{\,c^2(t)}\rho
+c^2(t)\Lambda (t)\qquad \quad \ 
\end{equation}
\begin{equation}
\label{e3}n^{\prime }+3nH-\psi =0
\end{equation}
where $n$ measures the particles number density, $\psi $ is the function
that measures the matter creation, $H=f^{\prime }/f$ represents the Hubble
parameter ($f$ is the scale factor that appears in the metrics), $p$ is the
thermostatic pressure, $\rho $ is energy density and $p_c$ is the pressure
that generates the matter creation.

The creation pressure $p_c$ depends on the function $\psi $. For adiabatic
matter creation this pressure takes the following form \cite{LI}: 
\begin{equation}
\label{w2}p_c=-\left[ \frac{\rho +p}{3nH}\psi \right] 
\end{equation}
The state equation used is the known expression 
\begin{equation}
\label{w3}p=\omega \rho 
\end{equation}
where $\omega =const.$ $\omega \in \left[ 0,1\right] $ physically realistic
equations, thus the energy-momentum tensor $T_{ij}$ verifies the energy
conditions.

It is necessary to know the exact form of the function $\psi $ , which is
determined from a more fundamental theory that involves quantum processes.
It is assumed that this function follows the law: 
\begin{equation}
\label{w5}\psi =3\beta nH 
\end{equation}
here we are following to Lima et al \cite{LI} (for other treatment \cite{DE}
while Prigogine et al \cite{PRI} follows this other law $\psi =\kappa H^2)$
where $\beta $ is a dimensionless constant (if $\beta =0$ then there is
matter creation since $\psi =0)$ presumably given by models of particles
physics of matter creation.

The conservation principle brings us to the following expression: 
\begin{equation}
\label{w4}\rho ^{\prime }+3(\omega +1)\rho \frac{f^{\prime }}f=(\omega
+1)\rho \frac \psi n 
\end{equation}

Integrating the equation (\ref{w4}) the following relationship between
energy density and the radius of the universe is obtained and even more
important the constant of integration necessary for our subsequent
calculations: 
\begin{equation}
\label{e4}\rho =A_{\omega ,\beta }f^{-3(\omega +1)(1-\beta )} 
\end{equation}
where $A_{\omega ,\beta }$ is the constant of integration that depends on
the state equation that is considered i.e. of the constant $\omega $ and of
the constant $\beta $ that measures the matter creation.

The effect of the bulk viscosity in the equations is shown replacing $p$ by $%
p-3\xi H$ where $\xi $ follow the law $\xi =\xi _0\rho ^\gamma $ (see \cite
{W},\cite{AB} and \cite{DE}). This last state equation, in our opinion, does
not verify the homogeneity principle for this reason it is modified by: 
\begin{equation}
\label{n2}\xi =k_\gamma \rho ^\gamma 
\end{equation}
where the constant $k_\gamma $ causes that this equation is indeed
dimensionally homogeneous for any value of $\gamma .$

The dimensional analysis followed needs to make the following distinctions:
it is necessary to know beforehand the set of fundamental quantities
together with that of unavoidable constants (in the nomenclature of
Barenblatt designated as governing parameters). In this case the only
fundamental quantity that appears in the model is the cosmic time $t$ as can
be easily deduced from the homogeneity and isotropy supposed for the model.
The unavoidable constants of the model are the constant of integration $%
A_{\omega ,\beta }$ that depends on the state equation $\omega $ and of the
mechanisms of matter creation $\beta $ and the constant $k_\gamma $ that
controls the influence of the viscosity in the model.

In a previous work \cite{T} the dimensional base was calculated for this
type of models, being this $B=\left\{ L,M,T,\theta \right\} $ where $\theta $
represents the dimension of the temperature. The dimensional equation of
each one of the governing parameters are:%
$$
\begin{array}{c}
\left[ t\right] =T\quad \left[ A_{\omega ,\beta }\right] =L^{3(\omega
+1)(1-\beta )-1}MT^{-2}\qquad \\ 
\left[ k_\gamma \right] =L^{\gamma -1}M^{1-\gamma }T^{2\gamma -1} 
\end{array}
$$
All the derived quantities or governed parameters in the nomenclature of
Batrenblatt will be calculated in function of these quantities (the
governing parameters), that is to say, in function of the cosmic time $t$
and of the two unavoidable constants $k_\gamma $ and $A_{\omega ,\beta }$
with respect to the dimensional base $B=\left\{ L,M,T,\theta \right\} .$

\section{{\bf {Solutions through D.A.}}}

Calculation will be made through dimensional analysis D.A. i.e. applying the
Pi Theorem, the variation of $G(t)$ in function of $t$, the speed of light $%
c(t),$ the Planck's constant $\hbar (t),$the radiation constant $a,$ the
charge of the electron $e(t),$ the mass of an elementary particle $m_i$\text{%
, }the variation of the cosmological ``constant'' $\Lambda (t),$ the energy
density $\rho (t),$ the matter density $\rho _m(t),$ the radius of the
universe $f(t),$ the temperature $\theta (t)$, the entropy $S(t)$ and the
entropy density $s(t),$the viscosity coefficient $\xi $$(t)$ and finally the
particle number density $n(t)\propto f^{-3}.$

The dimensional method brings us to (see \cite{T} and \cite{B}).

\subsection{{\bf Calculation of }$G(t)$}

As indicated above, will be accomplished calculation of the variation of $G$
applying the Pi theorem. The quantities considered are: $G=G(t,k_\gamma
,A_{\omega ,\beta }).$ with respect to the dimensional base $B=\left\{
L,M,T,\theta \right\} .$ We know that $\left[ G\right] =L^3M^{-1}T^{-2}$%
$$
\left( 
\begin{array}{ccccc}
& G & t & k_n & A_\omega \\ 
L & 3 & 0 & \gamma -1 & 3(\omega +1)(1-\beta )-1 \\ 
M & -1 & 0 & 1-\gamma & 1 \\ 
T & -2 & 1 & 2\gamma -1 & -2 
\end{array}
\right) 
$$
we obtain a single monomial that leads to the following expression for $G$%
\begin{equation}
\label{r1}G\propto A_{\omega ,\beta }^{\frac 2{3(\omega +1)(1-\beta )}}k_n^{
\frac{2+3(\omega +1)(1-\beta )}{3(\omega +1)(1-\beta )(\gamma -1)}%
}t^{-4-\left[ \frac{2+3(\omega +1)(1-\beta )}{3(\omega +1)(1-\beta )(\gamma
-1)}\right] } 
\end{equation}

\subsection{\bf Calculation of $c(t)$}

$c(t)=c(t,k_\gamma ,A_{\omega ,\beta })$ where $\left[ c\right]
=LT^{-1}\Longrightarrow $%
\begin{equation}
\label{r2}c(t)\propto A_{\omega ,\beta }^{\frac 1{3(\omega +1)(1-\beta
)}}k_n^{\frac 1{3(\omega +1)(1-\beta )(\gamma -1)}}t^{-1-\left[ \frac
1{3(\omega +1)(1-\beta )(\gamma -1)}\right] } 
\end{equation}

\subsection{\bf Calculation of the Planck's constant $\hbar (t):$}

$\hbar =\hbar (t,k_\gamma ,A_{\omega ,\beta })$ where the dimensional
equation is: $\left[ \hbar \right] =L^2MT^{-1}$ $\Longrightarrow $%
\begin{equation}
\label{r3}\hbar (t)\propto A_{\omega ,\beta }^{\frac 1{(\omega +1)(1-\beta
)}}k_n^{\frac{1-(\omega +1)(1-\beta )}{(\omega +1)(1-\beta )(\gamma -1)}}t^{
\frac{(\omega +1)(1-\beta )\left[ 1+(n-1)\right] -1}{(\omega +1)(1-\beta
)(\gamma -1)}} 
\end{equation}

\subsection{\bf Radiation constant $a(t):$}

$a=a(t,k_\gamma ,A_{\omega ,\beta },k_B)$ where $\left[ a\right]
=L^{-1}MT^{-2}\theta ^{-4}\Longrightarrow $%
\begin{equation}
\label{r4}k_B^{-4}a(t)\propto A_{\omega ,\beta }^{\frac{-4}{(\omega
+1)(1-\beta )}}k_n^{\frac{-4+3(\omega +1)(1-\beta )}{(\omega +1)(1-\beta
)(\gamma -1)}}t^{\frac{4-3(\omega +1)(1-\beta )}{(\omega +1)(1-\beta
)(\gamma -1)}} 
\end{equation}

\subsection{\bf Charge of the electron $e(t):$}

$e=e(t,k_\gamma ,A_{\omega ,\beta },\epsilon _0)$ where $\left[ e^2\epsilon
_0^{-1}\right] =L^3MT^{-2}$ $\Longrightarrow $%
\begin{equation}
\label{r5}e^2(t)\epsilon _0^{-1}\propto A_{\omega ,\beta }^{\frac 4{3(\omega
+1)(1-\beta )}}k_n^{\frac{4-3(\omega +1)(1-\beta )}{3(\omega +1)(1-\beta
)(\gamma -1)}}t^{\frac{-4+3(\omega +1)(1-\beta )}{3(\omega +1)(1-\beta
)(\gamma -1)}} 
\end{equation}

\subsection{M{\bf ass of an elementary particle $m_i(t):$}}

$m=m(t,k_\gamma ,A_{\omega ,\beta })$ where $\left[ m\right] =M$ $%
\Longrightarrow $%
\begin{equation}
\label{r6}m(t)\propto A_{\omega ,\beta }^{\frac 1{3(\omega +1)(1-\beta
)}}k_n^{\frac{1-3(\omega +1)(1-\beta )}{3(\omega +1)(1-\beta )(\gamma -1)}%
}t^{2-\left[ \frac{1-3(\omega +1)(1-\beta )}{3(\omega +1)(1-\beta )(\gamma
-1)}\right] } 
\end{equation}

\subsection{\bf Cosmological constant $\Lambda (t).$}

$\Lambda =\Lambda (t,k_\gamma ,A_{\omega ,\beta })$ where $\left[ \Lambda
\right] =L^{-2}$%
\begin{equation}
\label{r7}\Lambda (t)\propto A_{\omega ,\beta }^{\frac{-2}{3(\omega
+1)(1-\beta )}}k_n^{\frac{-2}{3(\omega +1)(1-\beta )(\gamma -1)}}t^{\frac
2{3(\omega +1)(1-\beta )(\gamma -1)}} 
\end{equation}
{\bf \ }

\subsection{\bf Calculation of energy density $\rho (t)$}

$\rho =\rho (t,k_\gamma ,A_{\omega ,\beta })$ with respect to the
dimensional base $B,$ where $\left[ \rho \right] =L^{-1}MT^{-2}$%
\begin{equation}
\label{r8}\rho \propto k_n^{\frac 1{1-\gamma }}t^{\frac 1{\gamma -1}} 
\end{equation}
it is observed that this relationship shows that energy density does not
depend either on the state equation $\omega $ or on the mechanisms on
creation of matter i.e. it does not depend on the constant $A_{\omega ,\beta
}$ solely on the viscosity of the fluid.

\subsection{\bf Matter density $\rho _m(t):$}

$\rho _m=\rho _m(t,k_\gamma ,A_{\omega ,\beta })$ where $\left[ \rho
_m\right] =ML^{-3}$ $\Longrightarrow $%
\begin{equation}
\label{r9}\rho _m(t)\propto A_{\omega ,\beta }^{\frac{-2}{3(\omega
+1)(1-\beta )}}k_n^{\frac{-2-3(\omega +1)(1-\beta )}{3(\omega +1)(1-\beta
)(\gamma -1)}}t^{2-\left[ \frac{-2-3(\omega +1)(1-\beta )}{3(\omega
+1)(1-\beta )(\gamma -1)}\right] } 
\end{equation}

\subsection{\bf Calculation of the radius of the universe $f(t).$}

$f=f(t,k_\gamma ,A_{\omega ,\beta })$ where $\left[ f\right]
=L\Longrightarrow $%
\begin{equation}
\label{r10}f\propto A_{\omega ,\beta }^{\frac 1{3(\omega +1)(1-\beta
)}}k_n^{\frac 1{3(\omega +1)(1-\beta )(\gamma -1)}}t^{\frac{-1}{3(\omega
+1)(1-\beta )(\gamma -1)}} 
\end{equation}
It can be observed that:%
$$
q=-\frac{f^{\prime \prime }f}{\left( f^{\prime }\right) ^2}=-1-3(\omega
+1)(1-\beta )(\gamma -1) 
$$
$$
H=\frac{f^{\prime }}f=-\left( \frac 1{3(\omega +1)(1-\beta )(\gamma
-1)}\right) \frac 1t 
$$

\subsection{\bf Calculation of the temperature $\theta (t).$}

$\theta =\theta (t,k_\gamma ,A_{\omega ,\beta ,}k_B)$ where $k_{B\text{ }}$
is the Bolztmann constant : $\left[ \theta \right] =\theta $ and $\left[
k_B\theta \right] =L^2MT^{-2}\Longrightarrow $%
\begin{equation}
\label{r11}k_B\theta \propto A_{\omega ,\beta }^{\frac 1{(\omega +1)(1-\beta
)}}k_n^{\frac{1-(\omega +1)(1-\beta )}{(\omega +1)(1-\beta )(\gamma -1)}%
}t^{-\left[ \frac{1-(\omega +1)(1-\beta )}{(\omega +1)(1-\beta )(\gamma -1)}%
\right] } 
\end{equation}

\subsection{{\bf Calculation of the entropy }$S(t):$}

$S=s(t,k_\gamma ,A_{\omega ,\beta },a)$ where $a$ is the radiation constant. 
$\left[ S\right] =L^2MT^{-2}\theta ^{-1}$%
\begin{equation}
\label{r12}S\propto A_{\omega ,\beta }^{\frac 1{(\omega +1)(1-\beta )}}k_n^{
\frac{1-\frac 34(\omega +1)(1-\beta )}{(\omega +1)(1-\beta )(\gamma -1)}%
}t^{-\left[ \frac{1-\frac 34(\omega +1)(1-\beta )}{(\omega +1)(1-\beta
)(\gamma -1)}\right] }a^{\frac 14} 
\end{equation}

\subsection{{\bf Calculation of the entropy density }$s(t):$}

$s=s(t,k_\gamma ,A_{\omega ,\beta },a)$ where $a$ is the radiation constant. 
$\left[ s\right] =L^{-1}MT^{-2}\theta ^{-1}$%
\begin{equation}
\label{r13}s\propto A_{\omega ,\beta }^0k_n^{\frac 3{4(\gamma
-1)}}t^{-\left[ \frac 3{4(\gamma -1)}\right] }a^{\frac 14} 
\end{equation}

\subsection{\bf Calculation of the viscosity coefficient $\xi $$(t):$}

$\xi $$=\xi (t,k_\gamma ,A_{\omega ,\beta })$ where $\left[ \xi \right]
=L^{-1}MT^{-1}$%
\begin{equation}
\label{r14}\xi \propto k_n^{\frac 1{1-\gamma }}t^{\frac{-\gamma }{\gamma -1}%
} 
\end{equation}

\subsection{\bf Particle number density $n(t).$}

$n=n(t,k_\gamma ,A_{\omega ,\beta })$ where $\left[ n\right] =L^{-3}$
obtaining: 
\begin{equation}
\label{r15}n(t)\propto A_{\omega ,\beta }^{\frac{-1}{(\omega +1)(1-\beta )}%
}k_n^{\frac{-1}{(\omega +1)(1-\beta )(\gamma -1)}}t^{\frac 1{(\omega
+1)(1-\beta )(\gamma -1)}} 
\end{equation}

\section{\bf Different cases.}

All the following cases can be calculated without difficulty. But as
indicated in the first section attention is centred only on those models
that follow the law $\xi =k_\gamma \rho ^{1/2}$ i.e. $\gamma =(1/2)$ wich
corresponds to models that are topologically equivalent to the classic FRW 
\cite{Z}. Two models with $\gamma $$=(1/2)$ are studied: one with $\omega
=1/3$ wich corresponds to a universe with radiation predominance and another
with $\omega =0$ corresponding to a universe with matter predominance.

\subsection{\bf Model with radiation predominance $\gamma $$=1/2$ and $%
\omega =1/3$.}

$$
\begin{array}{l}
G\propto A_\omega ^{\frac 1{2(1-\beta )}}k_n^{-2-\frac 1{(1-\beta
)}}t^{-2+\frac 1{(1-\beta )}}\quad  \\ 
c\propto A_\omega ^{\frac 1{4(1-\beta )}}k_n^{
\frac{-1}{2(1-\beta )}}t^{-1+\frac 1{2(1-\beta )}}\quad  \\ \hbar \propto
A_\omega ^{\frac 3{4(1-\beta )}}k_n^{2-\frac 3{2(1-\beta )}}t^{
\frac{1+2\beta }{2(1-\beta )}} \\ k_B^{-1/4}a\propto A_\omega ^{
\frac{-3}{(1-\beta )}}k_n^{\frac{6\beta }{1-\beta }}t^{\frac{6\beta }{\beta
-1}} \\ e^2\epsilon _0^{-1}\propto A_\omega ^{\frac 1{(1-\beta )}}k_n^{
\frac{-2\beta }{1-\beta }}t^{\frac{2\beta }{(1-\beta )}} \\ m\propto
A_\omega ^{
\frac{-1}{2(1-\beta )}}k_n^{2-\frac 1{2(1-\beta )}}t^{\frac 1{2(1-\beta
)}}\qquad  \\ \Lambda \propto A_\omega ^{-\frac 1{2(1-\beta )}}k_n^{\frac
1{(1-\beta )}}t^{-\frac 1{(1-\beta )}}\qquad 
\end{array}
$$
With these results it is proven that the relationship $G/c^2$ (general
covariance) remains constant without the need of imposing it as other
authors do \cite{P} and \cite{AVE}. 
\begin{equation}
\label{CO1}\frac G{c^2}=\frac{t^{-2+\frac 1{(1-\beta )}}}{t^{-2+\frac
1{(1-\beta )}}}=const.
\end{equation}
Likewise, it is observed that the fine structure constant remains constant
independently of the value of $\beta $%
$$
\alpha =\frac{e^2}{\epsilon _0c\hbar }=\frac{t^{\frac{2\beta }{(1-\beta )}}}{%
t^{-2+\frac 1{(1-\beta )}}t^{\frac{1+2\beta }{2(1-\beta )}}}=const. 
$$
that is to say, in this model a possible variation of the fine structure
constant $\alpha $ cannot be explained \cite{WEBB}. In the way in wich the
variation of the charge of the electron has been calculated it cannot be
discerned wheder $\epsilon _0$ is constant or not. Let us suppose that $%
\epsilon _0=const.$ (for an opposite point of view \cite{SUM} and the
appendix) therefore $e\propto t^{\frac \beta {(1-\beta )}}$ and from the
relationship $c^2=\frac 1{\epsilon _0\mu _0}$ is obtained $\mu _0\propto
c^{-2}\propto t^{2-\frac 1{(1-\beta )}}.$

If $\beta =0$ becomes (there is no matter creation) the following results
are found:%
$$
\begin{array}{c}
G\propto t^{-1}, 
\text{ }c\propto t^{-1/2},\text{ }\hbar \propto t^{1/2},\text{ }a=const. 
\text{ ,} \\ \text{ }e^2\epsilon _0^{-1}=const.\ ,\mu _0\propto t,\text{ }%
m_i\propto t^{1/2\text{ }}\text{ and }\Lambda \propto t^{-1} 
\end{array}
$$
The result of $G\propto t^{-1}$ is very well-known in the literature, the
value of $c\propto t^{-1/2}$ also has been obtained by Troiskii \cite{TR}
and Barrow \cite{BA} (in very different contexts), a similar result to $%
\hbar \propto t^{1/2}$ can be found in \cite{BLA},\cite{PEG},\cite{WES} and 
\cite{BAU} whereas a contrary point of view see \cite{SOL}, regarding the
constancy or not of the charge of the electron and of the fine structure
constant has been discussed (amongst other) by \cite{GAM}, $e^2\epsilon
_0^{-1}=const.$ has been obtained in particular if it is assumed that $%
\epsilon _0=const$ then is obtained that $e=const.$ and $\mu _0\propto t$,
with respect to $G\propto t^{-1}$ and $m_i\propto t^{1/2\text{ }}$ a similar
result to that is obtained by Hoyle\&Narlikar \cite{HOY} and Canuto et al 
\cite{CAN} a study on the implications of the variation of the masses can be
found in Mansfield et al \cite{MAN}.

With respect to rest of the quantities the same behavior is obtained as that
of Lima et at \cite{LI} except for the temperature $\theta $ and the
particle number density $n$.%
$$
\begin{array}{l}
\rho \propto k_n^2t^{-2}\quad \quad \quad \rho \propto t^{-2}\quad \\ 
k_B\theta \propto A_\omega ^{\frac 3{4(1-\beta )}}k_n^{2-\frac 3{2(1-\beta
)}}t^{-2+\frac 3{2(1-\beta )}} \\ 
f\propto A_\omega ^{\frac 1{4(1-\beta )}}k_n^{-\frac 1{2(1-\beta )}}t^{\frac
1{2(1-\beta )}} \\ 
a^{
\frac{-1}4}S\propto A_\omega ^{\frac 3{4(1-\beta )}}k_n^{-\frac{3\beta }{%
2(1-\beta )}}t^{\frac{3\beta }{2(1-\beta )}} \\ a^{
\frac{-1}4}s\propto A_\omega ^0k_n^{\frac 32}t^{-\frac 32} \\ \xi \propto
k_n^2t^{-1} \\ 
n\propto A_\omega ^{\frac{-3}{4(1-\beta )}}k_n^{\frac 3{2(1-\beta )}}t^{
\frac{-3}{2(1-\beta )}}\qquad 
\end{array}
$$
In the results, the temperature $\theta $ depends explicitly on $\beta $ on
such a way that $\beta <\frac 14$ so that the temperature of our universe
does not increase, on the contrary it cools down as it expands, with this
value of $\beta ,$ $q>\frac 12$ is obtained. The same happens with the value
obtained for $n,$ the result depends explicitly on the parameter $\beta $$.$
The rest of the quantities coincide with those obtained by Lima et al except
obviously for the quantity $\xi $ since their model describes a perfect
fluid. With these solutions our model does not have the horizon problem
posed by classic FRW since $ct=f.$

With respect to the thermodynamic behavior, the matter creation formulation
considered here is a clear consequence of the nonequilibrium thermodynamic
in presence of a gravitational field. We see that the $\beta $ parameter
works in the opposite sense to the expansion, that is, reducing the cooling
rate with respect to the case where there is no matter creation. A very
meaningful result is the fact that the spectrum of this radiation cannot be
distinguished from the usual blackbody spectrum at the present epoch (see 
\cite{LI}). Therefore models with adiabatic matter creation can be
compatible with the isotropy currently observed in the spectral distribution
of the background radiation. At the same timeit can be observed, that the
obtained model is clearly irreversible (classic FRW is reversible).

With these results the following law for the energy density%
$$
\rho =a\theta ^4 
$$
is recovered since the radiation constant $a$ also depends on $\beta $ and
obviously the law $\rho f^{4(1-\beta )}=const.$ is verified. It should be
pointed out, furthermore, that the variation of the entropy is due to the
variation of the radiation constant $a$. The energy follows the law%
$$
E=\hbar \nu \propto t^{\frac{-1+4\beta }{2(1-\beta )}}\qquad E=m_ic^2\propto
t^{\frac{-1+4\beta }{2(1-\beta )}} 
$$
since $E=k_B\theta .$

It should also be mentioned that all the important quantities of the classic
FRW models are recovered if $\beta =0$ is made \cite{T} :%
$$
\begin{array}{c}
f\propto t^{1/2},\ \rho \propto t^{-2},\ \theta \propto t^{-1/2},\ S=const.,
\\ 
\ s\propto t^{-3/2}\ ,n\propto t^{-3/2} 
\end{array}
$$
It is interesting to stick out that the model presented here may
significantly alter the predictions that the classic FRW make on the
abundance of elements. Such a result possibly limits the values that could
be taken by the $\beta $ parameter.

Finally and according to Prof. Alfonso-Faus's works (\cite{FAU}) the desire
is to show that the model exposed verifies, in addition to general
covariance principle (\ref{CO1}) the principles of Lorentz invariance, Mach
and Equivalence.

With regard to the Lorentz invariance it is seen that this is verified if
the relationship $v/c$ remains constant with time%
$$
\frac vc=const. 
$$
but this relationship is always constant since in our model all the speeds
vary following the law $v\propto t^{-1/2}$.

In relation to Mach's principle, this is verified if the following equality
is fulfilled%
$$
\frac{GMm}{f(t)}=mc^2 
$$
that establishes the equality between energy of a particle and gravitational
potential energy of the same. It can be proved without difficulty that even $%
\beta $ is verified$.$

Finally the Equivalence principle is verified if the following relationship%
$$
\frac{GM}{f^2(t)}\frac tc=const. 
$$
is maintained constant. It can be proven without difficulty that is verified
even for all values of $\beta $.

With regard to Planck's system (\cite{COU}), this now shows the following
behavior:%
$$
\begin{array}{l}
l_p=\left( 
\frac{G\hbar }{c^3}\right) ^{1/2}\approx f(t) \\ m_p=\left( 
\frac{c\hbar }G\right) ^{1/2}\approx f(t) \\ t_p=\left( \frac{G\hbar }{c^5}%
\right) ^{1/2}\approx t 
\end{array}
$$
with this behavior it is seen that this model does not has the designated
problem of Planck since the radius of the Universe $f(t)$ in the Planck's
era coincides with the length of Planck%
$$
f(t_p)\approx l_p 
$$
while energy density in the Planck's era coincides with energy density of
Planck%
$$
\rho (t_p)\approx \rho _p\approx t^{-2} 
$$
where $\rho _p=m_pc^2/l_p^3.$

\subsection{\bf Model with matter predominance $\gamma $$=1/2$ y $\omega =0:$%
}

$$
\begin{array}{l}
G\propto A_\omega ^{\frac 2{3(1-\beta )}}k_n^{-2-\frac 4{3(1-\beta
)}}t^{-2+\frac 4{3(1-\beta )}}\quad \\ 
c\propto A_\omega ^{\frac 1{3(1-\beta )}}k_n^{
\frac{-2}{3(1-\beta )}}t^{-1+\frac 2{3(1-\beta )}}\quad \\ \hbar \propto
A_\omega ^{\frac 1{(1-\beta )}}k_n^{
\frac{-2\beta }{(1-\beta )}}t^{\frac{1+\beta }{1-\beta }} \\ 
k_B^{-1/4}a\propto A_\omega ^{
\frac{-4}{(1-\beta )}}k_n^{\frac{2+6\beta }{1-\beta }}t^{\frac{2+6\beta }{%
\beta -1}} \\ e^2\epsilon _0^{-1}\propto A_\omega ^{\frac 4{3(1-\beta
)}}k_n^{1-\frac 8{3(1-\beta )}}t^{
\frac{2+6\beta }{3(1-\beta )}}\qquad \\ m_i\propto A_\omega ^{\frac
1{3(1-\beta )}}k_n^{2-\frac 2{3(1-\beta )}}t^{\frac 2{3(1-\beta )}} \\ 
\Lambda \propto A_\omega ^{-\frac 2{3(1-\beta )}}k_n^{\frac 4{3(1-\beta
)}}t^{-\frac 4{3(1-\beta )}}\qquad 
\end{array}
$$
With these results it is seen that exactly the same occurs as in the case of
radiation predominance i.e. that is, the relationship $G/c^{2\text{ }}$
(general covariance) remains constant and the fine structure constant also
remains constant in this case. It is easily proven that in the same way as
in the case of radiation predominance this model also fulfills the
principles stuied previously: Equivalence, Mach, Lorentz invariance.

If $\beta $$=0$ is made:%
$$
\begin{array}{c}
G\propto t^{-\frac 23},\ c\propto t^{-1/3},\ \hbar \propto t,\ a\propto
t^{-2}, \\ 
\ e\propto t^{\frac 13},\ m_i\propto t^{\frac 23},\ \Lambda \propto t^{-4/3} 
\end{array}
$$
is obtained: $c\propto t^{-1/3}$ is also obtained by Barrow \cite{BA} but
not by Troiskii \cite{TR}. We observe that with $\ \beta =0$ this time the
charge of the electron behaves as $e^2\epsilon _0^{-1}\propto t^{\frac 23}$
if $\epsilon _0=const.$ is considered, then $e\propto t^{\frac 13}$ while $%
\mu _0\propto t^{2/3}$. The radiation constant also varies $a\propto t^{-2}$%
. The masses continue varying in proportion to time while the gravitation
``constant'' varying as $G\propto t^{-\frac 23}.$ Finally it should be
pointed out that the Planck's constant varies direct proportion to time $%
\hbar \propto t.$

The rest of the quantities presents the following behavior: 
$$
\begin{array}{l}
\rho \propto k_n^2t^{-2} \\ 
\rho _m\propto A_\omega ^{
\frac{-2}{3(1-\beta )}}k_n^{2+\frac 4{3(1-\beta )}}t^{\frac{-4}{3(1-\beta )}%
}\quad \\ f\propto A_\omega ^{\frac 1{(1-\beta )}}k_n^{-\frac 2{3(1-\beta
)}}t^{\frac 2{3(1-\beta )}} \\ 
\xi \propto k_n^2t^{-1} 
\end{array}
$$
The law of temperatures obtained is: 
$$
k_B\theta \propto A_\omega ^{\frac 1{(1-\beta )}}k_n^{\frac{-2\beta }{%
(1-\beta )}}t^{\frac{2\beta }{(1-\beta )}} 
$$
as the temperature in the matter predominance era should be kept constant,
the only possibility is to make $\ \beta =0,$ i.e. during this era is no
creation of matter (in the case of not making $\ \beta =0$ our Universe
would be heated). With $\ \beta =0$ it is proven that energies are preserved 
$$
E=\hbar \nu =const.\qquad E=mc^2=const. 
$$
since $E=k_B\theta =const.$ This model with $\ \beta =0$ is very similar to
a FRW with matter predominance though here the problem of the horizon no
exits since $ct=f$ ,%
$$
\rho \propto t^{-2}\qquad \rho _m\propto t^{-4/3}\qquad f\propto t^{2/3} 
$$
With respect to the obtained result with the parameter $q=1/2$ it might seem
be in contradiction with the current observed values (acceleration of the
Universe) those which are based on the constancy of the luminosity of the
stars. However, in our case the luminosity varies in inverse proportionto to
time 
$$
L\propto \frac{GMm_pc}{\sigma _T}\approx t^{-1} 
$$
where $m_{p\text{ }}$ represents the proton mass and $\sigma _T$ is the
cross-section, i.e. the luminosity decreases with time. Sandage has
calculated the effect of the evolution on the luminosity of the galaxies.%
$$
\frac{L^{\prime }}L=10^{-9}/a\tilde no 
$$
In our case%
$$
\frac{L^{\prime }}L=t^{-1} 
$$
and for an age of the Universe about $10^{10}years$ does not disagree of our
result.\\

Before ending a reference should be made to Petit's work \cite{P}.

This author, in a very different context, gauge invariance, has studied the
variation of the physical constant, being one of the first to consider the
possible variation of the constant $c$ \cite{P1}. His results coincide with
ours for the case: $\left( \gamma =1/2,\ \omega =0,\ \beta =0\right) $ i.e.
an universe topologically equivalent to the classic FRW with MATTER
predominance and without creation of matter. However, Petit says to work
with an universe that describes the radiation era, said coincidence does not
exist with our work. However, we believe that his work is very correct in
the development, but in reality, he is describing an universe with matter
predominance by using in all his work the mass density (see in \cite{P}
equation number (32)). For this reason his model cannot verify the law $\rho
\propto f^{-4}$. If it is assumed (in our opinion) that his model describes
an universe with matter predominance, it is found that {\bf all} his results
coincide with the ours for the case above described i.e. $\left( \gamma
=1/2,\ \omega =0,\ \beta =0\right) ,$ these are:%
$$
\begin{array}{c}
G\propto t^{-\frac 23},\ c\propto t^{-1/3},\ \hbar \propto t,\ a\propto
t^{-2}, \\ 
\ e\propto t^{\frac 13},\ m_i\propto t^{\frac 23},\ \epsilon _0=const.,\ \mu
_0\propto t^{2/3} \\ 
f\propto t^{2/3}\qquad \rho _m\propto t^{-4/3} 
\end{array}
$$
\ 

\section{\bf Conclusions.}

The behavior of the ``constants'' within two specific models has been
calculated. In the first of the cases, a universe with radiation
predominance, it has been seen that the mechanisms of matter creation are
valid provided that $\ \beta <1/4$ since of the contrary our universe would
be heated as it expands. If we restrict ourselves to the case $\ \beta =0$
(non creation of matter) the solutions obtained are not discordant with
those already obtained by other authors. In this case it is found that the
radiation constant as well as the relationship $e^2\epsilon _0^{-1}$remain
constant if\ $\beta =0$ while the rest of the ``constants'' vary
independently of the value of\ $\beta $. The two models studied here verify
the general covariance principles $G/c^2$, Lorentz invariance $v/c=const.$,
Mach and Equivalence for all value of $\beta .$ It is also found that the
fine structure constant\ $\alpha $ remains constant since the ``constants''
that define it vary in such a way that the relationship remains constant$.$
To emphasize furthermore that with the variation of the constant of
radiation $a$ the relationship $\rho =a\theta ^4$\ \ is recovered for energy
density. Finally it should be commented that this model upon varying the
speed of the light does not have the problem of the horizon, being verified
the equality $ct=f$. It has also been possible to explain the so-called
Planck's problem as well as the entropy problem.

With respect to the model with matter predominance it is seen that in it
mechanisms of creation of matter cannot be considered, since if these are
taken in acount the temperature would increase instead of remaining constant
while the universe is expanded. With $\beta =0$ it is proven that energies
are preserved. In this case, the same as in the previous, we see that the
relationship $G/c^2$ remains constant the same as the fine structure
constant\ $\alpha $, but if\ $\gamma \neq 1/2$ these relationships do not
stay constant. Finally it should emphasized that in this case contrary to
what happend in the model of radiation predominance with\ $\beta =0$ the
``constants'' $a$ $\ $and $e$ vary.

\section{\bf Appendix. Behavior of the electromagnetic quantities.}

Following the observations of {\bf Prof. Alfonso-Faus} we explore other
possibilities from the results that it is obtained for the product
``inseparable'' $e^2\epsilon _0^{-1},$ this was showing the following
behavior depending on the era in the one which was calculated:%
$$
e^2\epsilon _0^{-1}=\left\{ 
\begin{array}{ccc}
const. &  & si 
\text{ }\omega =\frac 13 \\  &  &  \\ 
t^{\frac 23} &  & si\text{ }\omega =0 
\end{array}
\right. 
$$
It is studied below with more detail the different possibilities that we
this solution has.

\subsection{{\bf Case $e^2\epsilon _0^{-1}$ in the radiation era $\omega
=\frac 13$}}

In this case the relationship that it is obtained is:%
$$
e^2\epsilon _0^{-1}=const. 
$$
from this relationship it is deduced that:

\begin{enumerate}
\item  $e^2=\epsilon _0=const.$ Case envisaged above.

\item  $e^2=\epsilon _0$ being able to vary in any way.

\item  $e^2=\epsilon _0$ imposing the condition $\epsilon _0=\mu _0=\frac 1c$
. This condition is derived from the $TH\epsilon \mu $ formalism, devised by
Lightman and Lee (see \cite{LL}) and may be used to implement the Einstein's
Equivalence principle as presented by Will (see \cite{WILL}).\\From this
relationship we obtain furthermore that:%
$$
e^2=\epsilon _0=\mu _0=\frac 1c 
$$
we recall that in this case the speed of the light varies as: $c\propto
t^{-1/2}.$%
$$
e^2=\epsilon _0=\mu _0\propto t^{1/2} 
$$
being verified furthermore 
$$
e^2=\hbar \propto t^{1/2} 
$$
With respect to electrical and magnetic field these are behaved as:%
$$
E\propto t^{-5/4\text{ }}\text{ and \ }H\propto t^{-5/4\text{ }} 
$$
$$
H=E 
$$
the electromagnetic energy density behaved as 
$$
u_{EM}=\epsilon _0E^2+\mu _0H^2\propto t^{-2} 
$$
this result is coherent with the obtained for the radiation energy density $%
\rho =a\theta ^4\propto t^{-2}.$ With these results the constant of fine
structure continuous being constant and the Bohr radius behaved as the
radius of the Universe since the imposition of the condition $\epsilon
_0=\mu _0=\frac 1c$ not alters the behavior of the constant $\hbar .$%
$$
R_B=\frac{\hbar ^2\epsilon _0}{e^2m}\propto t^{1/2}\approx f(t) 
$$
\end{enumerate}

Respect to the result $\epsilon _0\propto t^{1/2}$ we observe that $\epsilon
_0\propto f(t)$ . This result already it was made clear by M\NEG oller (see 
\cite{MOLL}) and afterwards Landau and Lifshitz (see \cite{LAND}) arrived to
the same conclusion. This observation is the one which makes Sumner (see 
\cite{SUM}) since arrives to this coincidence result $\epsilon _0=\mu
_0\propto t^{1/2}$ that as we will see only we have in the case of radiation.

\subsection{{\bf Case $e^2\epsilon _0^{-1}$ in the matter era $\omega =0.$}}

In this case, the relationship that we had obtained was:%
$$
e^2\epsilon _0^{-1}\propto t^{2/3}. 
$$
from this relationship it is deduced that:

\begin{enumerate}
\item  $\epsilon _0=const.$ y $e\propto t^{1/3}.$ Case envisaged in the work.

\item  $e=const.$ y $\epsilon _0^{-1}\propto t^{2/3}.\Longleftrightarrow \mu
_0\propto t^{4/3}.$ It is remembered that in this case $c\propto t^{-1/3}.$
Concerning to the quantities $E$ and $H$ they show the next behaviour:%
$$
E\propto t^{-2/3\text{ }}\text{ and \ }H\propto t^{-5/3\text{ }} 
$$
while the electromagnetic energy density behaved as:%
$$
u_{EM}=\epsilon _0E^2+\mu _0H^2\propto t^{-2} 
$$
in this case the constant $\alpha $ also continuous being constant.

\item  Imposing the condition $\epsilon _0=\mu _0=\frac 1c$ ($TH\epsilon \mu 
$ formalism$)$ it is obtained the following results:%
$$
\epsilon _0=\mu _0=\frac 1c\text{ \quad /\quad }e^2\epsilon _0^{-1}\propto
t^{2/3} 
$$
we recall that in this case $c\propto t^{-1/3}.$ Then:%
$$
\epsilon _0=\mu _0\propto t^{1/3} 
$$
(the Sumner's results (see \cite{SUM}) no coincide with ours in this case)
while%
$$
e^2=\hbar \propto t 
$$
i.e. $e\propto t^{1/2}.$ Concerning to the quantities $E$ and $H$ show the
following behaviour:%
$$
E\propto t^{-7/6\text{ }}\text{ and \ }H\propto t^{-7/6\text{ }} 
$$
$$
H=E 
$$
while the Electromagnetic energy density behaved as:%
$$
u_{EM}=\epsilon _0E^2+\mu _0H^2\propto t^{-2} 
$$
In this case the constant $\alpha $ also continuous being constant in spite
of the fact that $\hbar \propto t$ i.e. our new results do not alter for
nothing the value of $\hbar $ already obtained and we verify the $TH\epsilon
\mu $ formalism. In this case the Bohr radius varies like the scale factor $%
f $ , the radius of the Universe%
$$
R_B=\frac{\hbar ^2\epsilon _0}{e^2m}\propto t^{2/3}\approx f(t) 
$$
while, Bohr total energy is maintained constant (result that not surprises
us since in the case of matter predominance (all) energies are preserved
while the moments not)%
$$
E_{TB}=\frac{me^4}{\epsilon _0^2\hbar ^2}=const. 
$$
\end{enumerate}

\subparagraph{\bf ACKNOWLEDGEMENTS:}

I wish to thank to Prof. Alfonso-Faus for suggestions and enlightening
discussions.\

\end{document}